\documentclass[usenatbib]{mn2e}
\usepackage[margin=0.75in,a4paper,portrait]{geometry}
\usepackage{amssymb}
\usepackage{graphicx}

\title[The O-C diagram of CS 1246]{Fortnightly Fluctuations in the O-C Diagram of CS 1246\thanks{Based on observations at the SOAR Telescope, a collaboration between CPNq-Brazil, NOAO, UNC, and MSU, and on observations collected at the PROMPT array via SKYNET.}}

\author[Barlow et al.]{B.N. Barlow$^{1}$\thanks{E-mail:bbarlow@physics.unc.edu}\thanks{Visiting astronomer, CTIO, National Optical Astronomy Observatories, which is operated by the Association of Universities for Research in Astronomy, Inc, under contract with the National Science Foundation}, B.H. Dunlap$^{1}\ddagger$, J.C. Clemens$^{1}$,
\newauthor D.E. Reichart$^{2}$, K.M. Ivarsen$^{2}$, A.P. LaCluyze$^{2}$, J.B. Haislip$^{2}$, M.C. Nysewander$^{3}$ \\
$^{1}$Department of Physics and Astronomy, University of North Carolina, Chapel Hill, NC 27599-3255, USA\\
$^{2}$SKYNET, Department of Physics and Astronomy, University of North Carolina, Chapel Hill, NC 27599-3255, USA\\
$^{3}$Alion Science \& Technology, 1000 Park Forty Plaza, Durham, NC 27713, USA}

\begin{document}

\date{Accepted 2011 March 3; received 2011 March 2; in original form 2010 September 28.}

\maketitle

\begin{abstract}
Dominated by a single, large-amplitude pulsation mode, the rapidly-pulsating hot subdwarf B star CS 1246 is a prime candidate for a long-term O-C diagram study.  We collected nearly 400 hours of photometry with the PROMPT telescopes over a time span of 14 months to begin looking for secular variations in the pulse timings.  Interestingly, the O-C diagram is dominated by a strong sinusoidal pattern with a period of 14.1 days and an amplitude of 10.7 light-seconds.  Underneath this sine wave is a secular trend implying a decrease in the 371.7-s pulsational period of $\dot{P}=-1.9 \times 10^{-11}$, which we attribute to the evolution of the star through the H-R diagram.  The sinusoidal variation could be produced by the presence of a low-mass companion with $m \sin i \simeq 0.12 M_{\sun}$ orbiting the subdwarf B star at a distance of 20 R$_{\sun}$.  Analysis of the combined light curve reveals the presence of a low-amplitude first harmonic to the main pulsation mode. 
\end{abstract}

\begin{keywords}
stars: individual: CS 1246 -- stars: oscillations -- stars: subdwarfs
\end{keywords}

\section{Hot Subdwarfs as Binary Systems}
\label{sec:intro}

Subluminous B (sdB) stars represent one of the least-understood stages of stellar evolution, yet they dominate surveys of faint blue objects and are found in almost all Galactic stellar populations.  They are the field counterparts of the Extreme Horizontal Branch (EHB) stars in globular clusters and lie in a region of the H-R diagram corresponding to the location of models with He-burning cores and extremely thin H envelopes \citep{heb86}. Understanding these stars better would illuminate the enigmatic ``second parameter" problem in globular cluster evolution, improve synthesis modeling of the UV upturn in giant elliptical galaxies, identify the evolutionary mechanisms responsible for guiding main sequence stars to this state, and constrain the physical properties of the dense plasmas present in these compact objects.     

Presumably, the sdB stars have been stripped of almost all their surface hydrogen after the red giant phase, but how and why this stripping occurs is still debated.   Some have proposed binary models in which the extra angular momentum provided by a companion manages to expel almost all of the hydrogen envelope \citep{han02,han03}.  Others have proposed that more modest amounts of angular momentum, as might be resident in a planet, could accomplish the same effect \citep{sok98}.  Still, others are able to draw hot subdwarf stars from their models without the assistance of a companion \citep{dcr96}. 

In any case, companions may be profoundly important for understanding the evolution of sdB stars, and placing constraints on the nature of these systems may shed light on their evolutionary histories.  Several studies have shown an unusually high binary fraction amongst the hot subdwarf stars \citep{max01,nap04b}.  Measurements of radial velocity (RV) shifts have led to the orbital parameters of approximately 85 sdB binaries (see Table A.1 from \citealt{gei11}).  The periods of such systems range from 0.07 to 28 days, with a peak between 0.5 d \& 1.0 d.  RV variability studies are most sensitive to systems with shorter periods and higher total masses, and as a result, few system parameters exist for longer-period binaries.  Only two systems with periods above 10 days have well-determined orbital parameters (PG 0850+170 \& PG 1619+522; \citealt{mor03a}).

The discovery of the first of the pulsating sdB (sdBV\footnote{Here we use the nomenclature proposed by \citet{nomenclature}.  All sdB pulsators are given the basename 'sdBV'; subscripts 'r' and 's' are added upon observations of rapid and slow pulsations, respectively.  For reference, the sdBV$_{r}$ stars have also been called V361 Hya and EC 14026 stars, while the sdBV$_{s}$ stars have the aliases PG 1716, V1093 Her, and 'Betsy' stars.}) stars \citep{kil97} in conjuction with the first sdBV star models \citep{cha96,cha97} opened the possibility of exploring their structure through a combination of spectral analysis and stellar seismology.  Multi-colour photometry and time-series spectroscopy techniques, for example, have revealed information about the total mass, envelope mass, and chemical stratification for a handful of pulsators (see \citet{ost09} for a review).  The pulsations may also provide the opportunity to detect companions by using phase measurements to look for oscillations in the arrival times of the pulses.  \citet{sil07} used this method successfully to detect a $3.2 M_{Jup}$ planet around the star V391 Peg (HS 2201+2610).  Their study also resulted in the first published measurement of $\dot{P}$ for a pulsating sdB star, constraining the evolutionary state of V391 Peg.

CS 1246, a recently-discovered sdBV$_{r}$ star \citep{bar10}, is a prime target for an extended O-C diagram study since it exhibits a single, large-amplitude pulsation mode. Using the PROMPT telescope array, we monitored the pulse timings for 14 months by collecting nearly 400 hours of time-series photometry.  The O-C diagram shows a sinusoidal oscillation superimposed on a parabola.  We interpret the quadratic term as a secular decrease in the pulsation period, while the sinusoidal oscillation implies the presence of a low-mass stellar companion, probably an M-dwarf or white dwarf.  Our study marks the second time $\dot{P}$ and the mass of a companion have been estimated for a hot subdwarf star using the O-C method.

\section{Photometric Data}
\label{sec:phot}

We began monitoring CS 1246 in 2009 using the 0.41-m Panchromatic Robotic Optical Monitoring and Polarimetry Telescopes (PROMPT) on Cerro Tololo to look for phase variations in the pulsations.  The PROMPT operate under the control of a prioritised, queue-scheduling system called SKYNET and are 100\% automated.  Priority levels are assigned to all observations in the queue, and, consequently, some of our photometry runs were interrupted by higher-priority targets such as gamma-ray bursts.  For additional details on the PROMPT and SKYNET, we refer the reader to \citet{rei05}.

\begin{figure}
\centering
	\includegraphics{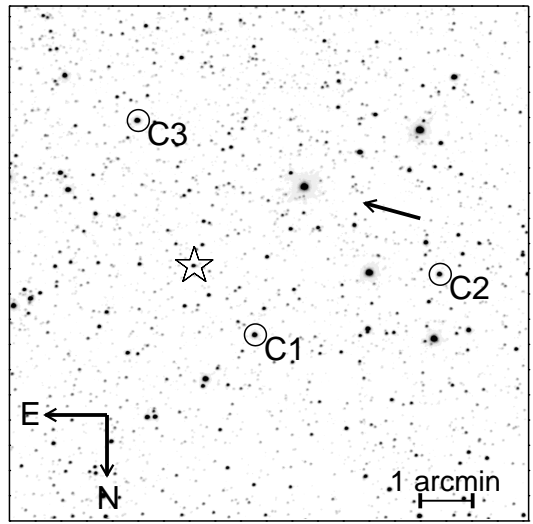}
\caption{Field image of CS 1246.  Image shown is a stack of 100 30-s Open-filter frames taken with PROMPT 3.  CS 1246 is marked with a star symbol, while the comparison stars used for the majority of the light curves are labelled C1-C3.  The arrow indicates the drift direction and magnitude of the field as it appears in the frames over three hours of tracking with PROMPT 3.}
\label{fig:field}
\end{figure}

Over the course of 14 months, we collected more than 33,000 individual frames representing 400 hours of time-series photometry.  The majority of these data were taken in a four-month span between January and May 2010 with PROMPT 3.  On a typical night, we obtained at least three hours of photometry with 30-s exposures and no filter; during these observations, our duty cycle was on average 83\%.  We also include in this work the photometry from 2009 presented in \citet{bar10} and a few light curves obtained with the Goodman Spectrograph \citep{cle04} on the 4.1-m SOAR telescope.  Appendix \ref{tab:phot_log} presents a detailed log of our photometric observations. 

\subsection{Reduction \& Analysis}
Images were bias-subtracted and flat-fielded in IRAF\footnote{IRAF is distributed by the National Optical Astronomy Observatories, which are operated by the Association of Universities for Research in Astronomy, Inc., under cooperative agreement with the National Science Foundation.} using standard procedures.  We then extracted our photometry using Antonio Kanaan's external IRAF package \textit{ccd\_hsp}.  We chose aperture radii that maximized the signal-to-noise ratio (S/N) in the nightly light curves and used sky annuli to estimate and subtract off counts from the sky.  As the field is densely packed with stars, we were careful to avoid contamination of the apertures and annuli from neighboring stars.  We removed sky transparency variations by dividing our light curves by those of constant comparison stars.  Such stars were chosen by selecting from those in the field that were photometrically constant in time, not contaminated by nearby stars, substantially brighter than CS 1246, and not saturated during a 30-s exposure.  Additionally, we avoided comparison stars that were likely to drift out of the frame during the PROMPT runs due to an approximately 0.4 arcmin hour$^{-1}$ tracking error in the mounts.  These selection criteria provided us with three dependable comparison stars, as shown in the field image in Figure \ref{fig:field}.  Lastly, we fit and normalised the light curves with parabolas to remove residual atmospheric extinction effects.  This process can remove real variations in the stellar flux on the order of the run lengths, making it difficult to detect lower-frequency \textit{g}-mode oscillations, if they exist.  All time stamps were converted to a barycentric Julian ephemeris date (BJED) using the WQED suite (v2.0; \citealt{tho09}), which employs the method of \citet{stu80}. 

\begin{figure}
\centering
	\includegraphics{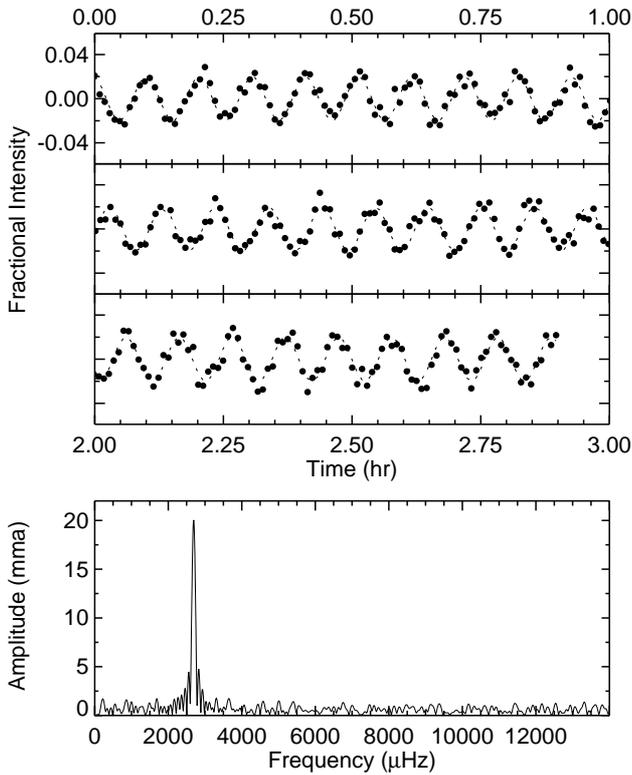}
\caption{(Top) Sample light curve of CS 1246 from 2010 Jan 18.  The data shown were obtained with PROMPT 3 over 2.9 hours using 30-s exposures without a filter.  (Bottom) Amplitude spectrum of the light curve, which is dominated by a 371.7-s signal.}
\label{fig:lc_sample}
\end{figure}

We analysed our light curves with a combination of Fourier analysis and least-squares fits of sine waves, using both \textit{Period04} \citep{len05} and MPFIT \citep{mar09}.  Initially, the period, phase, and amplitude were left as free parameters in the fits, but once we obtained a better determination of the period (see \S \ref{sec:O-C}), we refit all of the data with the period fixed.  Figure \ref{fig:lc_sample} shows a representative light curve from one of our photometry runs above its amplitude spectrum.

\subsection{Night-by-night results}

Each of the individual 89 light curves is dominated by the 371.7-s pulsation mode (hereafter $f_1$) reported by \citet{bar10} and shows no other significant signals in the \textit{p}-mode range of the amplitude spectrum (1600-14000 $\mu$Hz) after pre-whitening the signal.  Low-amplitude signals appear in some of the nightly Fourier transforms (FTs) at frequencies near the \textit{g}-mode regime, but we hesitate to claim these signals as real since 1/f noise and improperly removed extinction effects could easily account for those structures.  

\begin{figure}\vspace{3mm}
\centering
	\includegraphics{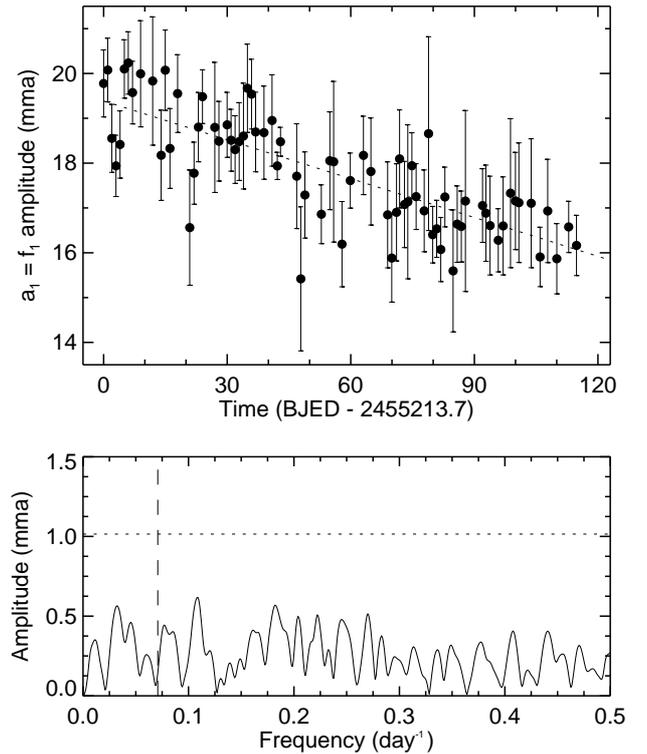}
\caption{(Top) Amplitudes of $f_1$ taken from the least-squares fits to the individual 2010 light curves.  The data show a decreasing trend of nearly 0.9 mma month$^{-1}$, which is marked by a dotted line.  (Bottom) Fourier transform of the nightly amplitudes with the linear trend removed.  The mean noise level is 0.25 mma.   The dashed vertical line marks the location of a 14.1-day oscillation, and the dotted horizontal line represents the amplitude at four times the mean noise level.}
	\label{fig:amplitudes}
\end{figure}

Our least-squares fits show the amplitude of $f_1$ decreased from 20 mma\footnote{Amplitudes are given in units of milli-modulation amplitude (mma), or parts-per-thousand. 10 mma corresponds to 1\%} in 2010 Jan to 16 mma in 2010 May at a rate of nearly 0.9 mma month$^{-1}$.  Figure \ref{fig:amplitudes} presents this trend above its FT.  We exclude the 2009 data from this plot since they were taken primarily through different passbands from the 2010 data and the measured amplitude of $f_1$ strongly depends on the observed wavelengths \citep{bar10}.  If the rate of decline remains linear, $f_1$ will have an amplitude below our detection limits (in nightly light curves) by 2011 Nov.  As we discuss later, there is no detectable periodic signal in the amplitudes.

The frequencies of $f_1$ in all the light curves agree within their error bars, which were typically around 1.2 $\mu$Hz.  To investigate smaller changes in the frequency as well as variations in the phase, we analysed our complete data set using the O-C technique, as discussed in the section that follows.

\section{The O-C Diagram}
\label{sec:O-C}

\begin{figure*}
\centering
	\includegraphics{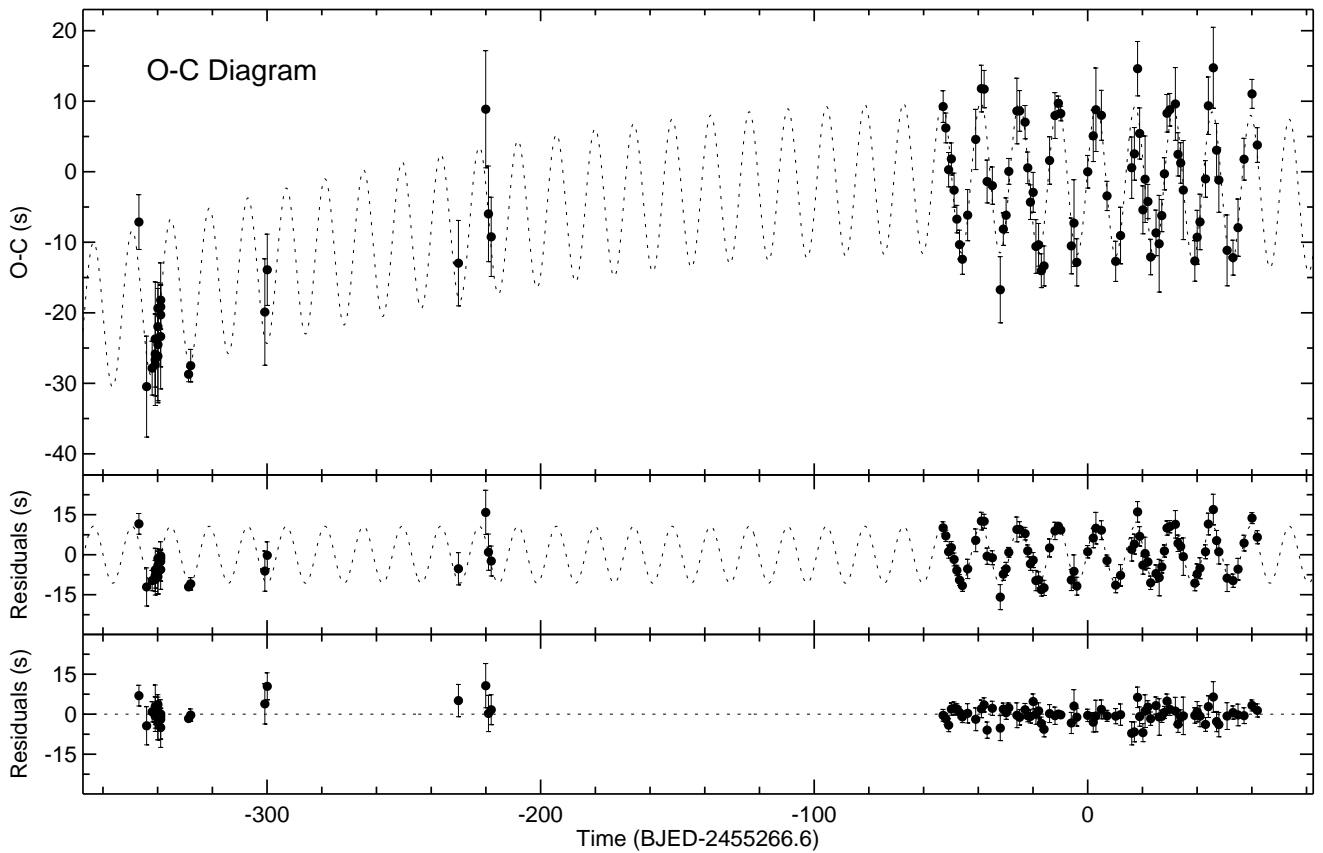}
\caption{(Top) The O-C diagram for CS 1246.  O-C values were computed using $f_1$ and a linear ephemeris.  The diagram is dominated by a strong sinusoidal pattern with a period of 14.1 days overlaid on a parabola.  (Middle) O-C points after removal of the quadratic term and (Bottom) after removal of both the parabola and sine wave.  The mean noise level in the pre-whitened diagram is 0.75s.}
\label{fig:O-C_full}
\end{figure*} 

We began our construction of the O-C diagram\footnote{For a review of the basic principles behind the O-C method, we refer the reader to \citet{kep93}.} by calculating a linear ephemeris for the times of maxima in the light curve ($C$) of the form \begin{equation}C=T_{c} + P_{c} E, \label{eqn:O-C_form}\end{equation} where $T_c$ is a reference time of maximum, $P_c$ the pulsation period, and $E$ the cycle number measured from $T_c$.   Since keeping track of $E$ correctly over an extended period of time requires an exceptionally accurate starting frequency, we used the period resulting from the least-squares fits to the combined 2010 light curve for $P_c$.  Our observing run on 2010 Mar 10 falls near the middle of this combined light curve, and, consequently, we employed the time of maximum determined from its least-squares fit for $T_c$.  The observed times of maxima ($O$) were taken from the least-squares fits to the individual light curves, and their corresponding cycle numbers were computed using equation (\ref{eqn:O-C_form}). 

\begin{figure}
\centering
	\includegraphics{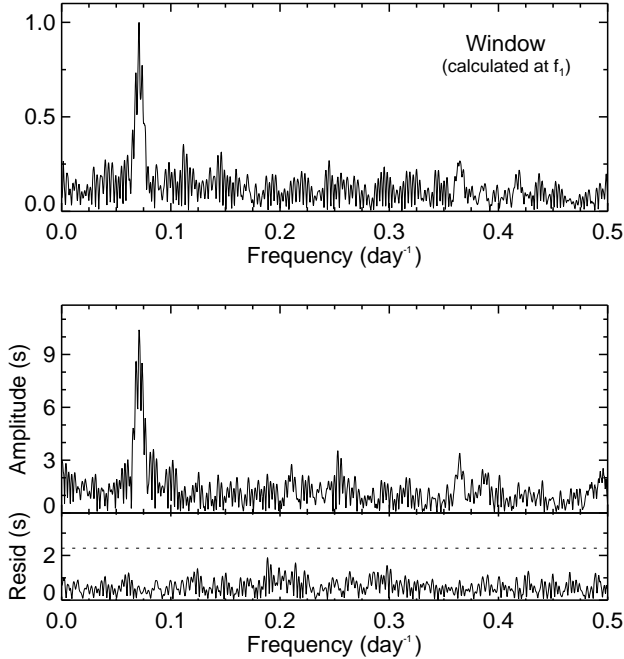}
\caption{Fourier transform of the O-C values after removal of the quadratic term (middle panel).  For reference, the window function for a sine wave sampled in the same manner as our data is also shown (top panel).  The Fourier transform of the data with the parabola and sine wave removed (bottom panel) has a mean noise level of 0.75 s.  The dashed line marks the 4-$\sigma$ level. }
\label{fig:O-C_ft}
\end{figure}

Figure \ref{fig:O-C_full} presents the O-C diagram created by subtracting the calculated from the observed times of maxima.  A sinusoidal oscillation dominates the overall structure.  The sinusoid is superimposed on a parabolic trend indicative of a secular change in the pulsation period.  The presence of the sinusoid is even more apparent in the FT of the O-C diagram with the quadratic term removed, as shown in Figure \ref{fig:O-C_ft}.  To quantify these structures, we performed a simultaneous fit to the O-C values including both parabolic and sinusoidal terms using the expression   \begin{equation}\label{eqn:O-C}O-C=\Delta T+\Delta P E+\frac{1}{2}P\dot{P}E^{2}+A \sin \left(\frac{2\pi E}{\Pi}+\phi \right).\end{equation}  We used the IDL routine MPFIT \citep{mar09}, which employs the Levenberg-Marquardt method, to perform a non-linear least-squares fit of equation (\ref{eqn:O-C}) to the data.  During this process, the points were weighted by their phase error bars as determined from the least-squares fits.  Inspecting both equations (\ref{eqn:O-C_form}) and (\ref{eqn:O-C}), one finds that the resulting values for $\Delta T$ and $\Delta P$ provide corrections to our initial estimates for the reference time of maximum and period from which we derive a final pulsational ephemeris for the times of maxima:  \begin{equation}\label{eqn:ephemeris} t_{max}=T_o + PE+\frac{1}{2}P\dot{P}E^{2}+A \sin \left(\frac{2\pi E}{\Pi}+\phi \right).\end{equation} Table \ref{tab:O-C_fit} displays the best-fit parameters for this ephemeris. 

The parabolic component of the fit indicates a secular decrease in the pulsational period on the order of 1 ms every 1.7 yrs.  We attribute this variation to structural changes in CS 1246 as it evolves, as we discuss further in \S \ref{sec:evolution}.  Removing this signal from the O-C data (Figure \ref{fig:O-C_full}, middle panel) more clearly reveals the two-week phase oscillation, which has a semi-amplitude of nearly 11s.  We point out that even the 2009 data, which are separated from the 2010 points by more than five months, phase well to this oscillation.  We find no additional phase variations after removing the parabolic and sinusoidal terms in either the O-C diagram (Figure \ref{fig:O-C_full}, bottom panel) or its FT (Figure \ref{fig:O-C_ft}, bottom panel). The mean noise level in the FT of the pre-whitened O-C diagram is 0.75 s.

\begin{table*}
\caption{Ephemeris parameters for the times of light maxima given in Eqn. \ref{eqn:ephemeris}}
\centering
\begin{tabular}{cllll}
\hline
Parameter & Value & Error & Units & Comments\\
\hline
T$_{o}$ & 2455266.604605 & $\pm$ 0.000004 & days & time of light maximum\\
$P$ & 371.69162& $\pm$ 0.00004 & s & $f_1$ pulsational period$^a$\\
 $\dot{P}$ & -1.9 $\times$ 10$^{-11}$ &$\pm$ 0.3$\times$ 10$^{-11}$ & $s\:s^{-1}$ & $f_1$ pulsational period change$^a$ \\
 A & 10.7 & $\pm$ 0.4 & s & fortnightly phase variation semi-amplitude\\
 $\Pi$ & 14.103 & $\pm$ 0.010 & days & fortnightly phase variation period\\
 $\phi$ & 0.023 & $\pm$ 0.007 & cycles & fortnightly phase variation phase$^a$\\
\hline
\multicolumn{5}{l}{\footnotesize{$^a$as measured at T$_o$}}\\
\end{tabular}
 \label{tab:O-C_fit}
\end{table*}

\newpage
\section{The Combined Light Curve}
\label{FT}

Three global variations associated with the pulsations of $f_1$ have been presented thus far:  an amplitude decrease, a two-week phase oscillation, and a decrease in the pulsation period.  Each of these phenomena should manifest itself in the FT of the combined light curve in a particular way.  For this reason and to look for lower-amplitude signals not detectable in the nightly light curves, we combined the light curves and computed the Fourier transform of the result.  The 2009 light curves were ignored in this analysis since they are few in number compared to the 2010 data and the large gaps they introduce to the combined light curve overly complicate the window function.  

\begin{figure*}
\centering
	\includegraphics{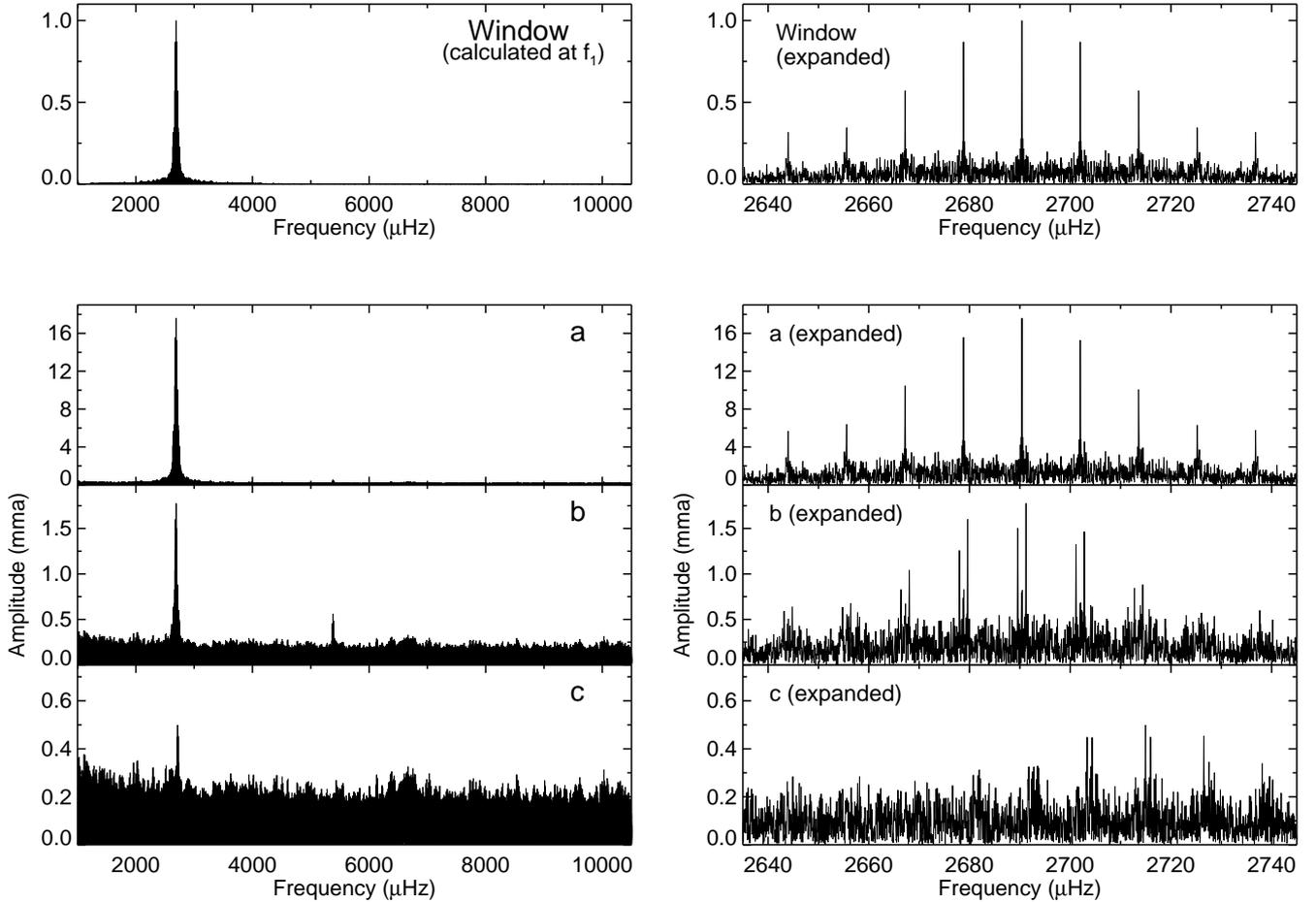}
\caption{(a) Amplitude spectrum of the combined 2010 light curve (left column) with expansions around $f_1$ (right column).  (b) Spectrum of the light curve pre-whitened by a sine wave with frequency $f_1$, which reveals a 0.82-$\mu$Hz splitting of $f_1$.  Note also the presence of the first harmonic to $f_1$.  (c) Amplitude spectrum of the data pre-whitened by the best fit of equation (\ref{eqn:lc_fit}). The mean noise level is around 0.074 mma.}
\label{fig:FT_lc}
\end{figure*}

Figure \ref{fig:FT_lc}a shows the amplitude spectrum of the complete light curve with an expanded view around $f_1$, which clearly dominates the spectrum.  After removing a sine wave of fixed frequency and amplitude from the light curve and re-computing the FT (Figure \ref{fig:FT_lc}b), one can see significant power at the location of the first harmonic of the main mode (5381 $\mu Hz$) with an amplitude near 0.6 mma.  The probability that random noise could produce this much signal by chance at exactly $2f_{1}$ is less than 10$^{-12}$; consequently, we regard this signal as real.  Evaluating the amplitude spectra of light curves combined month-by-month provides further support for the presence of the first harmonic.

Even more prominent than the first harmonic, the amplitude spectrum of the pre-whitened light curve reveals a substantial amount of residual structure surrounding the location of $f_1$.  The most significant of these signals is an apparent 0.82-$\mu Hz$ splitting of $f_1$ and its sidelobes, which is shown in the right half of Figure \ref{fig:FT_lc}b.  With an amplitude nearly 10\% that of the original mode, these components are split by exactly 1/14.1 day, corresponding to the period of the phase oscillation observed in the O-C diagram.  An additional splitting with an even tighter frequency separation and much smaller amplitude is also visible in the panel.

To test whether the above structures are consistent with the aforementioned three global variations associated with $f_1$, we modeled the complete light curve with the function   \begin{eqnarray} \label{eqn:lc_fit} I(t)=(a_1+\dot{a_1}t) \sin \left[ \frac{2\pi t}{P+\dot{P}t} + \phi_1 + \phi(t) \right]  \nonumber \\ + a_2 \sin \left[ \frac{2\pi t}{\frac{P}{2}+\frac{\dot{P}}{2}t} + \phi_2 + 2\phi(t) \right] ,  \end{eqnarray} \normalsize which includes a fundamental mode and first harmonic with linear period changes.  Additionally, the fundamental mode is given a linear amplitude variation.  Both terms contain a phase oscillation given by  \begin{equation}\phi(t) = \frac{2\pi A}{P+\dot{P}t} \sin \left( \frac{2\pi t}{\Pi} + \phi_\pi \right) . \end{equation} Fixing $\dot{P}$ to its value determined from the O-C diagram, we let all other parameters vary freely and used MPFIT to find the best-fitting values shown in Table \ref{tab:lc_fit}.   We note that all fitted parameters determined previously in this manuscript ($P$, $A$, $\Pi$, $\dot{a_1}$, $a_1$, $a_2$) using other methods (O-C diagram, night-by-night light curve fits) agree with those in the table to within the errors. Subtracting this fit from the data and re-calculating the FT (Figure \ref{fig:FT_lc}c) shows our model removed essentially all of the signals centred around $f_1$.  The only signal remaining in the amplitude spectrum after this process is a low-amplitude peak offset by 15 $\mu$Hz from $f_1$ at 2715 $\mu$Hz.   The frequency separation from $f_1$ implies an interaction timescale of 0.5 day.  The origin of this signal is unclear, but the probability that noise could produce it by chance is less than 10$^{-5}$. 

Other scenarios could also lead to the residual FT structure observed around $f_1$ in Figure \ref{fig:FT_lc}b, but they would have to do by conspiracy what our proposed model does naturally.  The equidistant triplets might be produced by (a) the splitting of $f_1$ by stellar rotation, (b) the presence of additional independent pulsation frequencies symmetrically disposed about the original, (c) one independent pulsation frequency and a combination mode at the difference between it and the first harmonic, and (d) amplitude variability of $f_1$.  None of these is appealing nor as effective as the model of equation (\ref{eqn:lc_fit}) at removing the residual power.  In particular, all of the alternative models predict an amplitude modulation of $f_1$ with a period of two weeks, which is not seen in the Fourier transform of the amplitude fits after the decreasing amplitude trend is removed (Figure \ref{fig:amplitudes}, bottom panel).

\begin{table*}
\caption{Best fit parameters to Eqn. (\ref{eqn:lc_fit})}
\centering
\begin{tabular}{cllll}
\hline
Parameter & Value & Error & Units & Comment\\
\hline
$a_1$  & 17.71 & $\pm$ $0.075$ & mma & $f_1$ amplitude$^a$\\
$\dot{a_1}$  & -0.0298 & $\pm$  $0.0027$ & mma day$^{-1}$ & $f_1$ amplitude change$^b$\\
$a_2$ & 0.5858 & $\pm$ $0.075$ & mma & 2$f_1$ amplitude$^{a,c}$\\
$P$ & $371.691646$ & $\pm$ $0.000033$ & s & $f_1$ period$^{a}$  \\
$\dot{P}$ & $-1.88$ &  $(fixed)$ & 10$^{-11}$ s s$^{-1}$ & $f_1$ period change$^d$\\
$A$ & $10.69$ & $\pm$ $0.35$ & s & $f_1$ phase oscillation semi-amplitude\\
$\Pi$ & $14.067$ & $\pm$ $0.032$ & day & $f_1$ phase oscillation period\\
$\phi_1$ & $0.85654$ & $\pm$ $0.00070$ & cycles & $f_1$ phase$^a$\\
$\phi_2$ & $0.435$ & $\pm$ $0.020$ & cycles & $f_2$ phase$^a$\\
$\phi_\pi$ & $0.8456$ & $\pm$ $0.0055$ & cycles & phase of the $f_1$ phase oscillation$^a$\\
\hline
\multicolumn{5}{l}{\footnotesize{$^a$ at t=0 corresponding to BJED 2455271.197436 }}\\
\multicolumn{5}{l}{\footnotesize{$^b$ from 2010 Jan to 2010 May}}\\
\multicolumn{5}{l}{\footnotesize{$^c$ with f$_2$ fixed to 2f$_1$}}\\
\multicolumn{5}{l}{\footnotesize{$^d$ fixed to value from Table \ref{tab:O-C_fit}}}
\end{tabular}
\label{tab:lc_fit}
\end{table*}

\section{Binary System Parameters}
\label{sec:system_parameters}

\begin{figure}
\centering 
	\includegraphics{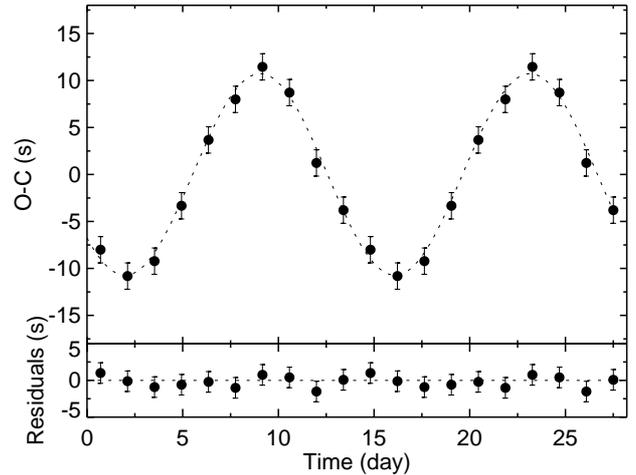}
\caption{(Top) Phase-folded O-C diagram with parabola removed.  The O-C points were folded on the fortnightly periodicity and averaged into ten bins.  The dotted line shows the circular orbit fit to the O-C diagram.  (Bottom) Residual O-C values after removal of a sinusoid fit.}
\label{fig:O-C_folded}
\end{figure}

Under the assumption that the O-C variations represent reflex motion due to an orbital companion, we have fitted the best orbital parameters.  Using the fits in Table \ref{tab:O-C_fit}, CS 1246 moves about a barycentre at least 11 light-seconds (4.7 R$_{\sun}$) away with a period of two weeks.  If the orbit is perfectly circular, the phase variation will be sinusoidal; otherwise, the shape will deviate from a sine wave.  As some subdwarf binaries show small eccentricities in their orbits \citep{ede06}, we decided to evaluate the eccentricity of the orbit by investigating both the phase-folded O-C curve and the Fourier transform thereof. Figure \ref{fig:O-C_folded} presents the phase-folded O-C diagram after removal of the parabolic trend.

The best-fitting orbital parameters to the folded O-C diagram yield an eccentricity of $\epsilon = 0.04 \pm 0.04$, which we take to be consistent with a circular orbit.  The absence of a prominent first harmonic (at 0.14 day$^{-1}$) in the FT of the O-C diagram (Figure \ref{fig:O-C_ft}, bottom panel), which would be expected for a considerably elliptical orbit, corroborates this conclusion.  Some degree of eccentricity might be expected since the circularization timescale for a binary with a two week orbital period exceeds 2 billion years \citep{tas92} but would not significantly alter the results of our mass determination.

Without information concerning the orbital motion of the unseen companion, we are limited to computing the mass function for the system: \begin{equation}f=\frac{m^{3}sin^{3}i}{(m+M)^{2}}=\frac{PK^{3}}{2\pi G}\end{equation}  The companion mass ($m$), sdB mass ($M$), and inclination angle ($i$) are free parameters, while the period ($P$) and RV semi-amplitude ($K$) of the sdB are taken from our fits to the O-C diagram.  To compute the mass of the companion, we must assume a mass for the sdB.  \citet{bar10} showed that under the assumption of a radial oscillation, a mass of 0.39 $^{+ 0.30}_{- 0.13}$ M$_{\sun}$ could be derived using the Baade-Wesselink method.  However, since the error bars are large and the radial nature of $f_1$ remains unproven, we use both this mass and the canonical sdB mass (0.47 $M_{\sun}$; \citealt{han02,han03}) to derive the system parameters shown in Table \ref{tab:system_parameters}.  

\begin{table}
\caption{System Parameters}
\centering
\scriptsize
\begin{tabular}{cllll}
\hline
Param & Value & Error & Units & Comment\\
\hline
$\Pi$  & 14.105 & $\pm$ 0.011 & days & orbital period\\
K & 16.6 & $\pm$ 0.6 & km s$^{-1}$ & RV semi-amplitude$^a$\\
$f$ & 0.0066 & $\pm$ 0.0007& M$_{\sun}$ & mass function\\
$a$  & 0.0910 & $\pm$  0.0003 & AU & separation distance$^b$\\
 & 0.0963 & $\pm$ 0.0003 & AU & separation distance$^c$\\
$m \sin i$ & 0.115 & $\pm$ 0.005 & $M_{\sun}$ & minimum companion mass$^{b,d}$\\
 & 0.129 & $\pm$ 0.005 &$M_{\sun}$ & minimum companion mass$^{c,d}$\\
\hline
\multicolumn{5}{l}{$^a$circular orbit approximation}\\
\multicolumn{5}{l}{$^b$assuming an sdB mass of 0.39 M$_{\sun}$}\\
\multicolumn{5}{l}{$^c$assuming the canonical sdB mass of 0.47 M$_{\sun}$}\\
\multicolumn{5}{l}{$^d$assumes no error bar on the sdB mass}
\end{tabular}
 \label{tab:system_parameters}
\end{table}

The minimum mass derived for the unseen companion is 0.12 $M_{\sun}$.  Further limitations could be placed on this mass by constraining the inclination angle of the system.  If the orbital plane is nearly edge-on ($i \simeq 90 \deg$), primary and secondary eclipses will occur when the O-C variations are close to their maximum and minimum values, respectively.  To look for an eclipse, we carefully monitored the light curve using three of the PROMPT telescopes on 2010 Jun 21 (UTC) at a time when equation (\ref{eqn:ephemeris}) predicts a primary eclipse of the sdB in the case of an edge-on orbital plane.  None of the light curves showed any sign of a signal loss.  Given the large separation distance, the absence of an eclipse only rules out inclination angles greater than $89 \deg$, which increases the minimum mass we report by a negligible amount.

Under the assumption of randomly distributed orbital plane orientations, one computes the probability of observing a system at an inclination angle $i$ less than or equal to $\theta$ as \begin{equation} Probability\: (i<\theta)=1-\cos \theta \end{equation}  From this we calculate a 96\% probability that the companion mass is less than  0.45  M$_{\sun}$, hence the object is most likely a low-mass white dwarf or late-type main sequence M-dwarf.  Optical spectra of CS 1246 \citep{bar10} show no clear signatures of a companion, but this result is consistent with both companion possibilities.  Optical reflection effects are often observed in sdB+dM systems, but even if the albedo of the M-dwarf were 1.0, the relatively large separation distance would result in a maximum flux increase less than 0.01 mma, a modulation well below our detection limits.  An infrared excess  would be expected for CS 1246 if the companion is a main sequence star with mass exceeding 0.45 M$_{\sun}$ \citep{lis05}.  Unfortunately, the presence of the Coalsack Nebula along our line of sight significantly reddens the system flux and complicates identification of an infrared excess from a companion.

\section{The Evolution of CS 1246}
\label{sec:evolution}

Clues to the current evolutionary state of CS 1246 may be found through consideration of the pulsational $\dot{P}$.  According to the models of \citet{cha02}, changes in the periods of acoustic modes in EHB stars are dominated by two primary phases.  During the first, a decrease in the surface gravity causes the periods to increase steadily as the sdB star evolves away from the zero-age extended horizontal branch (ZAEHB).  Model stars display positive values of $\dot{P}$ in this phase.  The second phase of evolution begins approximately 90 Myr after the ZAEHB when the density of thermonuclear fuel in the core begins to decrease dramatically.  Responding to the reduced energy production, the model stars contract, and the resulting increase in surface gravity leads to a decrease in acoustic mode periods and hence a negative value for $\dot{P}$.  This phase lasts until the onset of the post-EHB phase approximately 110 Myr after the ZAEHB, at which point the He in the core is completely exhausted.  Thereafter, He fusion may begin in the shell, the radius begins to increase, and the star heads towards the white dwarf cooling sequence after passing through the sdO regime of the H-R diagram.  Most helium-deficient sdO stars are believed to be the progeny of sdB stars evolving in this way.  
 
Our negative value of $\dot{P}$, interpreted as an evolutionary effect, implies CS 1246 is nearing the end of its life as an EHB star or has already done so.  The contraction resulting from the depletion of He in the core yields an increased surface gravity, which leads to a decrease in the pulsational period of -571 $\pm$ 85 s Myr$^{-1}$.  \citet{cha02} calculated values of $\dot{P}$ for sdB acoustic modes (their Appendix C) for representative models at different ages.  Comparing our $\dot{P}$ to all of their theoretical ones, we find our rate of change to be more rapid than all of their listed values but most closely matches that of a radial mode in a model with an age of 106.58 Myr and a hydrogen layer mass of 0.0042 M$_{\sun}$.  Our faster $\dot{P}$ could result from CS 1246 having a thicker hydrogen layer than any of their models, since \citet{cha02} show increasing the hydrogen layer mass leads to faster period changes.  It is also possible CS 1246 exists in an evolutionary epoch not represented in their table.  \citet{cha02} note that in advanced stages of EHB evolution, when the period decreases most rapidly, they used simple forward differencing to estimate theoretical $\dot{P}$ values instead of cubic spline fittings, which they used in earlier EHB stages when the periods are more stable.  A potential consequence of this differencing is the over- or under-estimation of period changes at these advanced evolutionary stages.  

Having stated these possibilities, we stress that our measurement of $\dot{P}$ is at most an \textit{upper limit} to the evolutionary rate of CS 1246.  Secular processes in sdB stars work on slower timescales than other processes affecting the phase or frequency of a mode, which in some cases drive the observed $\dot{P}$ to faster changes \citep{kaw10}.  Proper motion, for example, can contribute to the measured period changes in pulsating stars \citep{paj95}, but the expected contribution in our case is several orders of magnitude smaller than our observed $\dot{P}$ and inconsequential to our analysis.

\section{Discussion \& Outlook}

Using the O-C technique, we have discovered an apparent orbital reflex motion in CS 1246 and measured an upper limit of $\dot{P}$ for its main pulsation mode.  This marks the second time such quantities have been derived for an sdBV star using this method.  The first measurements were from \citet{sil07}.  The measured rate of period change, if secular, implies the star has either already exhausted the He in its core or is quickly approaching the post-EHB stage.  Models show the structure of sdB stars in this evolutionary state change dramatically fast, and it may even be possible to measure $\ddot{P}$ within a few years.  

\begin{figure}
\centering 
	\includegraphics{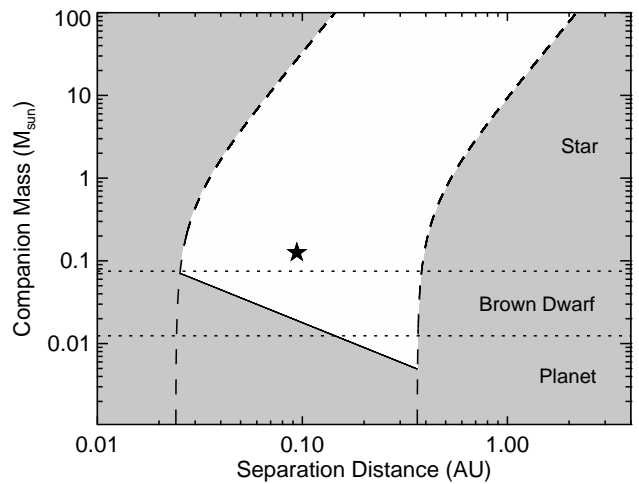}
\caption{Approximate sensitivity of our O-C diagram to companions around CS 1246.  The shaded region marks objects that are \textit{not} detectable using the O-C data shown in Figure \ref{fig:O-C_full}.  Companions falling within the white region could be detected in our data set, with certain restrictions on their orbital inclination angle.  For these calculations, we assume a mass of 0.47 M$_{\sun}$ for CS 1246.  The two dashed lines mark the combinations of companion masses and separation distances leading to orbital periods equal to the longest (right dashed line) and shortest (left dashed line) periods detectable using our O-C data, as computed using Kepler's third law.  The solid line shows the combinations of masses and distances resulting in reflex motions equal to our detection limits in the Fourier transform of the O-C diagram, under the assumption of an edge-on orbital plane.  Masses below this line induce reflex motion below our detection limits, regardless of their inclination angles.  The star symbol marks the location of the companion we detected (plotted using the minimum mass), and the two dotted horizontal lines show the boundaries between planet and brown dwarf masses (lower dotted line) and between brown dwarf and stellar masses (upper dotted line).}
\label{fig:sin_i}
\end{figure}

An analysis using Kepler's law shows the companion has a minimum mass near 0.12 M$_{\sun}$ and is most likely a  white dwarf or late-type main sequence star.  The separation distance implies the companion was potentially inside the envelope of the sdB progenitor during its red giant phase and might even be responsible for the ejection of the envelope (see \citealt{han02,han03}).  We cannot rule out the presence of additional bodies in the system, but we can limit the possible combinations of companion masses and separation distances using the detection limits from the Fourier transform of our O-C data.  The shaded region in Figure \ref{fig:sin_i} shows combinations of masses and separation distances leading to phase wobbles that are \textit{undetectable} using our current data set.  Our ability to detect a companion with orbital parameters inside the white region depends on the inclination angle of its orbit.  The lower solid line shows the combinations of system parameters resulting in reflex motions equal to our detection limits, under the assumption of an edge-on orbital plane; the line moves up with increasing inclination angle.  The region is bounded on the left and right by dashed lines corresponding to the minimum and maximum orbital periods we can detect given the sampling rate and baseline of our O-C data; thus, we could expand the area of the white region by obtaining phase measurements more frequently or by monitoring the system for a longer period of time.  Regardless of the sampling, additional phase measurements will lower our detection limits and move the lower boundary (solid line) to smaller masses.  As a reference, the star symbol shows the location of the single companion we detect around CS 1246 (plotted with its minimum mass).  If the observed phase oscillation is the result of a binary companion, CS 1246 would show the second-lowest RV semi-amplitude and the third-longest period of the $\sim$85 sdB binary systems for which orbital parameters are known (Table A.1, \citealt{gei11}).

More statistics are needed to draw conclusions about the formation channels leading to sdB stars from the orbital parameters of binary systems.  Searches for radial velocity variations, as are being carried out by the MUCHFUSS survey \citep{gei11}, should uncover a wide variety of sdB binary systems, although they are somewhat biased towards higher-mass systems with shorter orbital periods.  Pulsating sdB stars may also be used to find unseen companions using the O-C diagram, as shown in this work and by \citet{sil07}, and the EXOTIME project \citep{sch10} is currently building up O-C diagrams for several sdBV stars to look for oscillations in the pulse timings.  Although both of these surveys will be effective at uncovering binary companions to sdB stars, their sampling rates differ significantly from the one presented in this study, and, consequently, they will not be sensitive to the same parameter space shown in Figure \ref{fig:sin_i}.   

Our result provides an important link between the two aforementioned methods for detecting unseen companions since an observable radial velocity variation should accompany our fortnightly phase oscillation.  The O-C diagram is a significant tool used to discover and characterize the nature of unseen companions, including extrasolar planets, but no corroborative RV variations have been measured to date.  The O-C oscillation of CS 1246 implies a two-week velocity variation with a semi-amplitude well within the range of medium-to-high resolution spectrographs.  Consequently, we are planning follow-up observations to measure the RV variations.  If the expected RV variation is observed, it will be the first independent confirmation that it is possible to detect stellar and planetary companions using the O-C technique and sdB pulsations as the clock. 

\section*{Acknowledgments}

We acknowledge the support of the National Science Foundation, under award AST-0707381, and are grateful to the Abraham Goodman family for providing the financial support that made the spectrograph possible.  BNB acknowledges additional support from a GAANN fellowship, Department of Education award number P200A090135.  We thank the Delaware Asteroseismic Research Center for providing the S8612 filter used in these studies.  We also recognize the observational support provided by the SOAR telescope operators Alberto Pasten, Patricio Ugarte, Sergio Pizarro, and Daniel Maturana.  The SOAR telescope is operated by the Association of Universities for Research in Astronomy, Inc., under a cooperative agreement between the CNPq, Brazil, the National Observatory for Optical Astronomy (NOAO), the University of North Carolina, and Michigan State University, USA.

 \appendix
\section{Observations Log}
 
 \begin{table*}
 \label{tab:phot_log}
 \tiny 
 \caption{Observations Log}
 \begin{tabular}{ccccccccccccccc}
 \hline
Date & Start Time & T$_{exp}$ & T$_{cycle}$ & Length & Filter & Scope$^{a}$ & & Date & Start Time & T$_{exp}$ & T$_{cycle}$ & Length & Filter & Scope$^{a}$\\
(UTC) & (UTC) & (s) & (s) & (hr) & & & &(UTC) & (UTC) & (s) & (s) & (hr) & & \\
\hline
 
                   2009-03-29                              &                      05:13:09                              &    40                              &    46                              & 1.23                              &                                V                              &                         P4
 & \hspace{6mm} & 
                      2010-02-21                              &                      02:00:07                              &    30                              &    36                              & 2.91                              &                             Open                              &                         P3
\\
                      2009-04-01                              &                      00:51:02                              &    40                              &    46                              & 1.40                              &                                V                              &                         P5
 & \hspace{6mm} & 
                      2010-02-22                              &                      02:07:14                              &    30                              &    36                              & 2.41                              &                             Open                              &                         P3
\\
                      2009-04-03                              &                      01:16:16                              &    20                              &    26                              & 1.17                              &                             Open                              &                         P4
 & \hspace{6mm} & 
                      2010-02-23                              &                      01:52:20                              &    31                              &    37                              & 2.48                              &                             Open                              &                         P3
\\
                      2009-04-03                              &                             23:59:03                              &    80                              &    86                              & 9.43                              &                               u'                              &                         P3
 & \hspace{6mm} & 
                      2010-02-25                              &                      02:27:30                              &    30                              &    36                              & 2.40                              &                             Open                              &                         P3
\\
                      2009-04-04                              &                             00:57:03                              &    40                              &    52                              & 9.57                              &                               g'                              &                         P2
 & \hspace{6mm} & 
                      2010-02-27                              &                      01:37:09                              &    30                              &    36                              & 3.11                              &                             Open                              &                         P3
\\
                      2009-04-04                              &                             00:01:48                              &    40                              &    47                              & 9.57                              &                               r'                              &                         P4
 & \hspace{6mm} & 
                      2010-02-28                              &                      08:04:58                              &    20                              &    26                              & 1.78                              &                            S8612                              &                         SOAR
\\
                      2009-04-04                              &                             00:01:52                              &    40                              &    47                              & 8.66                              &                               i'                              &                         P5
 & \hspace{6mm} & 
                      2010-03-01                              &                      02:37:10                              &    20                              &    26                              & 6.79                              &                              B                              &                         SOAR
\\
                      2009-04-04                              &                             23:57:56                              &    80                              &    86                              & 8.61                              &                               u'                              &                         P3
 & \hspace{6mm} & 
                      2010-03-05                              &                      01:13:28                              &    30                              &    36                              & 2.41                              &                             Open                              &                         P3
\\
                      2009-04-04                              &                             23:57:06                              &    40                              &    52                              & 9.31                              &                               g'                              &                         P2
 & \hspace{6mm} & 
                      2010-03-06                              &                      01:09:45                              &    30                              &    36                              & 1.61                              &                             Open                              &                         P3
\\
                      2009-04-04                              &                             23:57:42                              &    40                              &    47                              & 9.29                              &                               r'                              &                         P4
 & \hspace{6mm} & 
                      2010-03-07                              &                      01:06:47                              &    30                              &    36                              & 2.92                              &                             Open                              &                         P3
\\
                      2009-04-04                              &                             23:57:39                              &    40                              &    47                              & 9.43                              &                               i'                              &                         P5
 & \hspace{6mm} & 
                      2010-03-11                              &                      00:44:23                              &    30                              &    36                              & 3.20                              &                             Open                              &                         P3
\\
                      2009-04-05                              &                             23:56:42                              &    80                              &    86                              & 9.70                              &                               u'                              &                         P3
 & \hspace{6mm} & 
                      2010-03-13                              &                      02:09:46                              &    30                              &    36                              & 1.64                              &                             Open                              &                         P3
\\
                      2009-04-05                              &                             23:55:47                              &    40                              &    52                              & 9.69                              &                               g'                              &                         P2
 & \hspace{6mm} & 
                      2010-03-14                              &                             00:32:23                              &    30                              &    36                              & 3.19                              &                               Open                              &                         P3
\\
                      2009-04-05                              &                             23:56:36                              &    40                              &    47                              & 9.68                              &                               r'                              &                         P4
 & \hspace{6mm} & 
                      2010-03-16                              &                      01:00:50                              &    30                              &    36                              & 2.59                              &                             Open                              &                         P3
\\
                      2009-04-05                              &                             23:56:32                              &    40                              &    47                              & 9.69                              &                               i'                              &                         P5
 & \hspace{6mm} & 
                      2010-03-18                              &                      01:21:58                              &    30                              &    36                              & 4.37                              &                             Open                              &                         P3
\\
                      2009-04-16                              &                      09:07:06                              &    10                              &    16                              & 1.10                              &                              R                              &                         SOAR
 & \hspace{6mm} & 
                      2010-03-21                              &                      06:02:18                              &    30                              &    36                              & 3.41                              &                             Open                              &                         P3
\\
                      2009-04-17                              &                      01:00:22                              &    80                              &    86                              & 6.14                              &                              350-620nm$^b$                              &                         SOAR
 & \hspace{6mm} & 
                      2010-03-23                              &                      01:20:24                              &    30                              &    36                              & 2.56                              &                             Open                              &                         P3
\\
                      2009-05-14                              &                      05:09:12                              &    20                              &    26                              & 3.08                              &                                V                              &                         P5
 & \hspace{6mm} & 
                      2010-03-27                              &                      03:00:52                              &    30                              &    36                              & 2.14                              &                             Open                              &                         P3
\\
                      2009-05-15                              &                             00:57:56                              &    20                              &    26                              & 5.45                              &                               V                              &                         P5
 & \hspace{6mm} & 
                      2010-03-28                              &                      02:51:33                              &    30                              &    36                              & 1.71                              &                             Open                              &                         P3
\\
                      2009-07-23                              &                             23:36:05                              &    40                              &    46                              & 3.25                              &                               V                              &                         P4
 & \hspace{6mm} & 
                      2010-03-29                              &                      06:30:50                              &    30                              &    36                              & 2.99                              &                             Open                              &                         P3
\\
                      2009-08-02                              &                      23:35:20                              &    40                              &    46                              & 2.67                              &                           rprime                              &                         P3
 & \hspace{6mm} & 
                      2010-03-30                              &                      00:28:42                              &    30                              &    36                              & 3.00                              &                             Open                              &                         P3
\\
                      2009-08-04                              &                      00:00:41                              &    40                              &    46                              & 1.48                              &                           rprime                              &                         P3
 & \hspace{6mm} & 
                      2010-03-31                              &                      05:06:29                              &    30                              &    36                              & 2.99                              &                             Open                              &                         P3
\\
                      2009-08-05                              &                      00:00:48                              &    40                              &    46                              & 2.12                              &                           rprime                              &                         P3
 & \hspace{6mm} & 
                      2010-04-01                              &                      00:29:16                              &    30                              &    36                              & 1.98                              &                             Open                              &                         P3
\\
                      2010-01-17                              &                      04:18:07                              &    30                              &    36                              & 3.25                              &                             Open                              &                         P3
 & \hspace{6mm} & 
                      2010-04-02                              &                      00:17:15                              &    30                              &    36                              & 3.97                              &                             Open                              &                         P3
\\
                      2010-01-18                              &                      04:13:15                              &    30                              &    36                              & 2.90                              &                             Open                              &                         P3
 & \hspace{6mm} & 
                      2010-04-03                              &                      00:50:48                              &    30                              &    36                              & 3.49                              &                             Open                              &                         P3
\\
                      2010-01-19                              &                      04:09:46                              &    30                              &    36                              & 3.88                              &                             Open                              &                         P3
 & \hspace{6mm} & 
                      2010-04-04                              &                      23:58:13                              &    30                              &    36                              & 3.08                              &                             Open                              &                         P3
\\
                      2010-01-20                              &                      04:05:29                              &    30                              &    36                              & 3.94                              &                             Open                              &                         P3
 & \hspace{6mm} & 
                      2010-04-06                              &                             01:26:05                              &    30                              &    36                              & 7.95                              &                               Open                              &                         P3
\\
                      2010-01-21                              &                      04:01:50                              &    30                              &    36                              & 3.90                              &                             Open                              &                         P3
 & \hspace{6mm} & 
                      2010-04-07                              &                             00:46:28                              &    30                              &    36                              & 5.28                              &                               Open                              &                         P3
\\
                      2010-01-22                              &                      04:52:00                              &    33                              &    39                              & 2.91                              &                             Open                              &                         P3
 & \hspace{6mm} & 
                      2010-04-08                              &                             00:41:50                              &    30                              &    36                              & 4.67                              &                               Open                              &                         P3
\\
                      2010-01-23                              &                      03:53:38                              &    25                              &    31                              & 3.38                              &                             Open                              &                         P3
 & \hspace{6mm} & 
                      2010-04-09                              &                      00:42:05                              &    30                              &    36                              & 2.98                              &                             Open                              &                         P3
\\
                      2010-01-24                              &                      04:39:12                              &    30                              &    36                              & 3.42                              &                             Open                              &                         P3
 & \hspace{6mm} & 
                      2010-04-10                              &                      00:38:21                              &    30                              &    36                              & 3.59                              &                             Open                              &                         P3
\\
                      2010-01-26                              &                      03:42:18                              &    30                              &    36                              & 1.48                              &                             Open                              &                         P3
 & \hspace{6mm} & 
                      2010-04-12                              &                             00:12:12                              &    30                              &    36                              & 5.56                              &                               Open                              &                         P3
\\
                      2010-01-29                              &                      01:51:06                              &    30                              &    36                              & 3.74                              &                             Open                              &                         P3
 & \hspace{6mm} & 
                      2010-04-13                              &                             00:07:47                              &    30                              &    36                              & 4.99                              &                               Open                              &                         P3
\\
                      2010-01-31                              &                      03:22:26                              &    30                              &    36                              & 3.93                              &                             Open                              &                         P3
 & \hspace{6mm} & 
                      2010-04-14                              &                             00:03:33                              &    30                              &    36                              & 3.70                              &                               Open                              &                         P3
\\
                      2010-02-01                              &                      03:18:25                              &    30                              &    36                              & 4.96                              &                             Open                              &                         P3
 & \hspace{6mm} & 
                      2010-04-15                              &                      00:03:48                              &    30                              &    36                              & 1.90                              &                             Open                              &                         P3
\\
                      2010-02-02                              &                      06:41:19                              &    30                              &    36                              & 1.93                              &                             Open                              &                         P3
 & \hspace{6mm} & 
                      2010-04-19                              &                      03:51:05                              &    30                              &    36                              & 4.29                              &                             Open                              &                         P3
\\
                      2010-02-04                              &                      03:06:53                              &    30                              &    36                              & 4.59                              &                             Open                              &                         P3
 & \hspace{6mm} & 
                      2010-04-19                              &                             23:49:39                              &    30                              &    36                              & 5.41                              &                               Open                              &                         P3
\\
                      2010-02-07                              &                      01:56:24                              &    30                              &    36                              & 2.69                              &                             Open                              &                         P3
 & \hspace{6mm} & 
                      2010-04-20                              &                             23:39:39                              &    30                              &    36                              & 7.40                              &                               Open                              &                         P3
\\
                      2010-02-08                              &                             04:19:12                              &    30                              &    36                              & 5.38                              &                               Open                              &                         P3
 & \hspace{6mm} & 
                      2010-04-22                              &                             23:37:12                              &    30                              &    36                              & 6.63                              &                               Open                              &                         P3
\\
                      2010-02-09                              &                             04:00:47                              &    30                              &    36                              & 3.84                              &                               Open                              &                         P3
 & \hspace{6mm} & 
                      2010-04-24                              &                      03:47:37                              &    30                              &    36                              & 1.84                              &                             Open                              &                         P3
\\
                      2010-02-10                              &                      04:01:02                              &    30                              &    36                              & 4.04                              &                             Open                              &                         P3
 & \hspace{6mm} & 
                      2010-04-25                              &                      22:58:30                              &    30                              &    36                              & 2.82                              &                             Open                              &                         P3
\\
                      2010-02-13                              &                      03:49:45                              &    30                              &    36                              & 2.55                              &                             Open                              &                         P3
 & \hspace{6mm} & 
                      2010-04-27                              &                      03:31:02                              &    30                              &    36                              & 3.66                              &                             Open                              &                         P3
\\
                      2010-02-14                              &                      03:59:15                              &    30                              &    36                              & 2.33                              &                             Open                              &                         P3
 & \hspace{6mm} & 
                      2010-04-27                              &                      23:33:40                              &    30                              &    36                              & 2.50                              &                             Open                              &                         P3
\\
                      2010-02-16                              &                      03:51:20                              &    30                              &    36                              & 2.33                              &                             Open                              &                         P3
 & \hspace{6mm} & 
                      2010-04-30                              &                      23:16:38                              &    30                              &    36                              & 3.58                              &                             Open                              &                         P3
\\
                      2010-02-17                              &                      02:49:32                              &    30                              &    36                              & 3.00                              &                             Open                              &                         P3
 & \hspace{6mm} & 
                      2010-05-03                              &                      02:56:35                              &    30                              &    36                              & 4.16                              &                             Open                              &                         P3
\\
                      2010-02-18                              &                      02:11:52                              &    30                              &    36                              & 3.11                              &                             Open                              &                         P3
 & \hspace{6mm} & 
                      2010-05-04                              &                      23:28:40                              &    30                              &    36                              & 3.11                              &                             Open                              &                         P3
\\
                      2010-02-19                              &                      02:08:17                              &    30                              &    36                              & 3.23                              &                             Open                              &                         P3
 & \hspace{6mm} & 
                      2010-05-07                              &                      04:31:20                              &    30                              &    36                              & 2.92                              &                             Open                              &                         P3
\\
                      2010-02-20                              &                      02:58:25                              &    30                              &    36                              & 2.21                              &                             Open                              &                         P3
 & \hspace{6mm} & 
                      2010-05-10                              &                       23:23:16                               &    30                              &    36                              & 4.15                              &                               Open                              &                         P3
\\

\hline
\multicolumn{15}{l}{$^a$``P" = PROMPT }\\
\multicolumn{15}{l}{$^b$light curve produced from time-series spectra over the wavelength range 3500-6200 \AA }\\
\end{tabular}
\end{table*}


\section{O-C Points} 
\begin{table*}
\caption{Observed times of light maxima and corresponding O-C values}
\scriptsize
\centering
\begin{tabular}{ccccccc}
\hline
Time of maximum & Error & O-C & & Time of maximum & Error & O-C\\
(BJED-2450000) & (s) & (s) & & (BJED-2450000) & (s) & (s)\\
\hline

  4919.752522 & $\pm$ 3.9 &  -7.1 & &     5248.652306 & $\pm$ 3.0 & -10.4\\
     4922.574358 & $\pm$ 7.2 & -30.5 & &     5249.646023 & $\pm$ 2.4 & -14.0\\
     4924.587722 & $\pm$ 3.8 & -27.8 & &     5250.635488 & $\pm$ 2.8 & -13.4\\
     4925.701949 & $\pm$ 6.5 & -26.7 & &     5252.661896 & $\pm$ 3.4 &   1.6\\
     4925.714875 & $\pm$ 1.8 & -25.8 & &     5254.640884 & $\pm$ 3.3 &   8.0\\
     4925.714857 & $\pm$ 3.2 & -27.3 & &     5255.879877 & $\pm$ 1.0 &   9.7\\
     4925.732098 & $\pm$ 8.1 & -23.7 & &     5256.757465 & $\pm$ 1.0 &   8.3\\
     4926.682811 & $\pm$ 6.6 & -26.2 & &     5260.607526 & $\pm$ 4.0 & -10.5\\
     4926.700089 & $\pm$ 3.7 & -19.4 & &     5261.584115 & $\pm$ 6.2 &  -7.3\\
     4926.700058 & $\pm$ 2.1 & -22.0 & &     5262.616527 & $\pm$ 3.3 & -12.8\\
     4926.704334 & $\pm$ 8.0 & -24.5 & &     5266.604618 & $\pm$ 2.3 &   0.0\\
     4927.698121 & $\pm$ 7.4 & -23.4 & &     5268.630913 & $\pm$ 3.6 &   5.1\\
     4927.711063 & $\pm$ 3.1 & -19.2 & &     5269.598903 & $\pm$ 5.9 &   8.8\\
     4927.711074 & $\pm$ 2.1 & -18.2 & &     5271.603620 & $\pm$ 3.5 &   8.0\\
     4927.698156 & $\pm$ 7.4 & -20.3 & &     5273.655536 & $\pm$ 2.1 &  -3.4\\
     4937.910976 & $\pm$ 1.1 & -28.7 & &     5276.830294 & $\pm$ 2.8 & -12.7\\
     4938.681036 & $\pm$ 2.3 & -27.5 & &     5278.615661 & $\pm$ 4.0 &  -9.0\\
     4965.787953 & $\pm$ 7.5 & -19.9 & &     5282.681150 & $\pm$ 4.4 &   0.6\\
     4966.665615 & $\pm$ 5.1 & -13.9 & &     5283.662026 & $\pm$ 3.7 &   2.5\\
     5036.555718 & $\pm$ 6.1 & -13.0 & &     5284.845213 & $\pm$ 3.9 &  14.6\\
     5046.545184 & $\pm$ 8.3 &   8.9 & &     5285.593652 & $\pm$ 3.6 &   5.4\\
     5047.538762 & $\pm$ 6.8 &  -6.0 & &     5286.785176 & $\pm$ 3.4 &  -5.4\\
     5048.549697 & $\pm$ 5.6 &  -9.2 & &     5287.572491 & $\pm$ 6.1 &  -1.1\\
     5213.750512 & $\pm$ 2.2 &   9.2 & &     5288.604931 & $\pm$ 2.4 &  -4.2\\
     5214.739933 & $\pm$ 2.1 &   6.2 & &     5289.615806 & $\pm$ 2.5 & -12.1\\
     5215.759436 & $\pm$ 2.4 &   0.3 & &     5291.573250 & $\pm$ 3.2 &  -8.7\\
     5216.757515 & $\pm$ 2.3 &   1.8 & &     5292.726167 & $\pm$ 6.9 & -10.2\\
     5217.755525 & $\pm$ 2.4 &  -2.6 & &     5293.676951 & $\pm$ 2.3 &  -6.2\\
     5218.770747 & $\pm$ 1.9 &  -6.7 & &     5294.636362 & $\pm$ 2.3 &  -0.3\\
     5219.738651 & $\pm$ 2.1 & -10.3 & &     5295.600108 & $\pm$ 2.7 &   8.3\\
     5220.771104 & $\pm$ 2.1 & -12.4 & &     5296.611080 & $\pm$ 2.3 &   8.8\\
     5222.715675 & $\pm$ 3.6 &  -6.2 & &     5298.654533 & $\pm$ 5.2 &   9.6\\
     5225.662660 & $\pm$ 4.3 &   4.6 & &     5299.618096 & $\pm$ 3.1 &   2.5\\
     5227.727697 & $\pm$ 3.3 &  11.8 & &     5300.594632 & $\pm$ 2.9 &   1.2\\
     5228.747268 & $\pm$ 2.6 &  11.7 & &     5301.549629 & $\pm$ 7.0 &  -2.6\\
     5229.826914 & $\pm$ 3.0 &  -1.4 & &     5305.761159 & $\pm$ 2.9 & -12.6\\
     5231.732689 & $\pm$ 2.6 &  -2.0 & &     5306.617291 & $\pm$ 3.8 &  -9.3\\
     5234.644963 & $\pm$ 4.7 & -16.7 & &     5307.649793 & $\pm$ 3.9 &  -7.1\\
     5235.754975 & $\pm$ 2.3 &  -8.1 & &     5309.663194 & $\pm$ 2.6 &  -1.0\\
     5236.765965 & $\pm$ 2.5 &  -6.2 & &     5310.708696 & $\pm$ 4.1 &   9.4\\
     5237.759795 & $\pm$ 1.8 &   0.0 & &     5312.524198 & $\pm$ 5.7 &  14.7\\
     5240.719663 & $\pm$ 4.7 &   8.6 & &     5313.711413 & $\pm$ 3.8 &   3.1\\
     5241.722024 & $\pm$ 2.9 &   8.6 & &     5314.541647 & $\pm$ 4.6 &  -1.2\\
     5243.718128 & $\pm$ 2.4 &   7.0 & &     5317.557226 & $\pm$ 5.0 & -11.1\\
     5244.686000 & $\pm$ 2.2 &   0.6 & &     5319.721109 & $\pm$ 2.5 & -12.2\\
     5245.662495 & $\pm$ 2.5 &  -4.3 & &     5321.562408 & $\pm$ 4.1 &  -7.9\\
     5246.664874 & $\pm$ 2.8 &  -2.9 & &     5323.760837 & $\pm$ 3.0 &   1.8\\
     5247.675753 & $\pm$ 3.8 & -10.6 & &     5326.669086 & $\pm$ 2.0 &  11.0\\

\hline
\end{tabular}
\end{table*}

 \bsp

\end{document}